\documentclass[pra, twocolumn]{revtex4}
\usepackage[ansinew]{inputenc}
\usepackage{graphicx}
\usepackage{amsmath}
\usepackage{amsthm}
\usepackage{bm}
\usepackage{layout}
\usepackage{float}
\usepackage{cancel}
\usepackage{amsfonts}
\usepackage{amssymb}
\usepackage{txfonts}%
\usepackage{nameref}
\usepackage{hyperref}
\newcommand\id{\leavevmode\hbox{\small1\kern-3.3pt\normalsize1}}

\newcommand{\tr}{\mbox{Tr}}

\begin{document}

	\title{Time-reparameterisation invariant quantum evolution law: the lack of absolute time does not imply a stationary global state}

	\author{Ognyan Oreshkov$^1$ and Denis Bouvy$^2$}

	\affiliation{$^1$QuIC, Ecole polytechnique de Bruxelles, CP 165, Universit\'{e} libre de Bruxelles, 1050 Brussels, Belgium.\\
$^2$Faculté des Sciences, Universit\'{e} libre de Bruxelles, 1050 Brussels, Belgium.}

\begin{abstract}

We challenge the common belief that if there is no absolute time parameter in physics, a quantum system described without reference to an external clock can be assumed to be in a stationary state of its Hamiltonian. We present a time-reparameterisation invariant quantum evolution law, which for a given initial condition predicts the same trajectory in state space as the Schrödinger equation, except that for nontrivial trajectories it does not predict the speed at which the trajectory is traversed. The solutions of this evolution law are all time-reparameterised solutions of the Schrödinger equation. We show how the predictions of the Schrödinger equation are recovered relative to an internal clock in this framework. In contrast to the Page-Wootters formalism or Dirac's quantisation of the Hamiltonian constraint, here the global sate is not stationary. We discuss the assumptions leading to the common conclusion that the state can be taken stationary and suggest that they need to be revisited.

\end{abstract}

\maketitle

\section{Introduction}

In quantum theory, the time evolution of the state $|\psi(t)\rangle$ of an isolated quantum system with Hamiltonian $\hat{H}$ is governed by the Schrödinger equation,
 \begin{gather}
\partial_t |\psi(t)\rangle = -i \hat{H}|\psi(t)\rangle, \label{Schrodinger}
\end{gather}
where $t$ is the time parameter (we use units where $\hbar =1$).

It is often stated that, since $t$ is an external parameter in this equation, if there is no absolute time parameter in physics, a system in the absence of an external clock, such as the universe, can only be in a stationary state of its Hamiltonian. The nontrivial evolution of subsystems within such a universe is then supposed to arise from the correlations between these subsystems and other subsystems that serve as clocks. This is the idea behind the Page-Wootters proposal for ``evolution without evolution'' \cite{PW}. The proposal assumes that the Hamiltonian of the universe is approximately of the form 
\begin{equation}
\hat{H}_{CS} = \hat{H}_C\otimes \hat{\id}_S + \hat{\id}_C\otimes \hat{H}_S, \label{PWHam}
\end{equation}
 where $\hat{H}_C$ is a Hamiltonian acting on a clock subsystem with Hilbert space $\mathcal{H}_C$ characterised by an orthonormal basis $|t\rangle_C$, $t \in (-\infty, \infty)$, $\langle t|t'\rangle = \delta(t-t')$, that approximately \cite{footnote1} obeys $e^{-i\int_{0}^{T} d\tau \hat{H}_C} |t\rangle_C = |t+T\rangle_C$, and $\hat{H}_S$ is a Hamiltonian on the rest of the universe (the `system') with Hilbert space $\mathcal{H_S}$. It then shows that any stationary density matrix $\hat{\rho}_{CS}$ on the full Hilbert space $\mathcal{H}_{CS} = \mathcal{H}_C\otimes \mathcal{H}_S$, 
 \begin{equation}
 [\hat{H}_{CS}, \hat{\rho}_{CS}]= 0, \label{PWcondition}
\end{equation}
 has the property that the family of conditional states $\hat{\rho}_S(t) \equiv \langle t|_C\hat{\rho}_{CS}|t \rangle_{C}/ \tr_S{\langle t|_C\hat{\rho}_{CS}|t \rangle_{C}}$ respects the Schrödinger equation for density matrices (or von Neumann-Liouville equation), 
 \begin{gather}
\partial_t \hat{\rho}_S(t) = -i [\hat{H}_S, \hat{\rho}(t)_S].
 \end{gather}
In other words, the Schrödinger evolution of the system $S$ is found encoded in the correlations between $C$ and $S$. 

For a pure state $|\psi\rangle_{CS}$ satisfying 
\begin{gather}
\hat{H}_{CS}|\psi\rangle_{CS} = 0,
\end{gather}
one analogously finds that $|\psi(t)\rangle_S \equiv \langle t |_C|\psi\rangle_{CS} / ||\langle t |_C|\psi\rangle_{CS} ||_S$, where $|| \cdot ||_S$ denotes the norm of a vector in $\mathcal{H}_S$,  satisfies
 \begin{gather}
\partial_t |\psi(t)\rangle_S = -i \hat{H}_S|\psi(t)\rangle_S.\label{localpureSchr}
\end{gather}

However, the interpretation of this mathematical result as a mechanism by means of which the time evolution that we observe appears in a universe whose global state is stationary raises many questions. If in an everyday laboratory we engineer a clock system $C'$ with Hamiltonian $\hat{H}_{C'}$ with the properties described above, and another system $S'$ with Hamiltonian $\hat{H}_{S'}$, such that the joint Hamiltonian is $\hat{H}_{C'S'} = \hat{H}_{C'}\otimes \hat{\id}_{S'} + \hat{\id}_{C'}\otimes \hat{H}_{S'}$, and we prepare the joint system in a stationary state $|\psi\rangle_{C'S'}$, say $\hat{H}_{C'S'}|\psi\rangle_{C'S'} = 0$, shall we say that $S'$ undergoes time evolution relative to $C'$ within this non-evolving state? If not, how are $C'$ and $S'$ different from $C$ and $S$? If so, what exactly justifies interpreting this as time evolution, as opposed to, say, spatial correlations? If the same state $|\psi\rangle_{C'S'}$ is prepared but the Hamiltonian $\hat{H}_{C'S'}$ is made to vanish (in which case the state would also be stationary), shall we still interpret the correlations in this state as internal time evolution? If so, given that there is now nothing that singles out the basis $|t\rangle_{C'}$, should we interpret correlations in any basis as time evolution? If not, how does the presence of a Hamiltonian that does not affect the state trigger the purported internal evolution relative to the right basis? And how is the duration of that evolution to be reconciled with the arbitrarily short time for which the state may exist in the laboratory? 

A similar situation arises in the context of Dirac's quantisation of systems with gauge freedom \cite{Dirac}. If we have a classical system with a time-reparameterisation invariant Lagrangian, in the Hamiltonian picture one typically finds that the Hamiltonian is, up to an arbitrary time-dependent factor, proportional to a constraint $H$. This constraint is a function of the phase-space variables that is identically zero when one expresses the canonical momenta in terms of the generalised coordinates and their derivatives, but it generates nontrivial evolution through the Poisson brackets. The Hamiltonian constraint is first-class, meaning that it has vanishing Poisson brackets with all other constraints in the theory. In Dirac's quantisation procedure (which first eliminates constraints that are not first-class by introducing the Dirac brackets), the phase-space variables are promoted to operators with the brackets replaced by commutators with prefactor $-i/\hbar$. The physical states are then defined as those that are annihilated by the operators of first-class constraints. For the Hamiltonian constraint, this yields 
\begin{gather}
\hat{H}|\psi\rangle = 0.\label{DiracHamConstraint}
\end{gather}
In the quantisation of general relativity, this corresponds to the Wheeler-DeWitt equation \cite{DeWitt}. 

An instructive illustration of this method is given by the case of a time-reparameterisation invariant system obtained from a system without time-reparameterisation invariance by promoting the original time parameter $t$ to a dynamical variable $t(\tau)$ (intuitively, one may think of $t$ as the position of the hand of a physical clock) that depends on the newly introduced time parameter $\tau$. With respect to $\tau$, the action becomes reparameterisation invariant, and the Hamiltonian constraint reads \cite{Henneaux} $H = p_C + H_S =0$, where $p_C$ is the momentum conjugate to $t$ and $H_S$ is the Hamiltonian of the unparameterised system. Quantised by Dirac's procedure, this yields the following condition for the physical states: 
\begin{gather} 
\left(\hat{p}_C\otimes \hat{\id}_S+ \hat{\id}_C\otimes \hat{H}_S\right)|\psi\rangle_{CS} = 0.
\end{gather} 
Note that, in contrast to $\hat{H}_C$ in \eqref{PWHam}, here $\hat{p}_C$ is unbounded from below, and so is the total Hamiltonian. As a generator, $\hat{p}_C$ generates the ideal translation of the time shown by the clock $C$ that $\hat{H}_C$ generates approximately. Defining $|\psi(t)\rangle_S \equiv \langle t |_C|\psi\rangle_{CS}$ (here the normalisation can be absorbed in that of $|\psi\rangle_{CS}$), one finds \cite{Henneaux} that $|\psi(t)\rangle_S$ respects Eq. \eqref{localpureSchr}, while the general solution for $|\psi\rangle_{CS}$ has the form \cite{Giovanetti} 
 \begin{gather}
|\psi\rangle_{CS}  = \int dt |t\rangle_C |\psi(t)\rangle_S.\label{constraintsolution}
\end{gather}
Although the starting assumptions here are different from those of Page and Wootters, the final picture is similar---we have a stationary vector in the Hilbert space that encodes Schrödinger evolution in the correlations between the system and a `clock' (see Ref. \cite{Colosi} for a similar analysis in the case of naturally time-reparameterisation invariant systems). The interpretation of this result raises similar conceptual questions. 

In this paper, we show that the absence of an absolute time parameter, which for a system described without reference to an external clock renders the time parameter gauge, does not imply a stationary state in quantum theory. Starting from the Schrödinger equation, we construct an underdetermined evolution law that is time-reparametrisation invariant. For a given initial condition, it predicts the same trajectory in state space as the Schrödinger equation, except that it does not predict the speed at which the trajectory is traversed for nontrivial trajectories, and hence the time parameterisation of the trajectory. The solutions of this law are all possible time-reparameterised solutions of the Schrödinger equation. We then show how the standard Schrödinger equation for a subsystem arises relative to a clock if the Hamiltonian has the same form \eqref{PWHam} as in the Page-Wootters proposal. But in contrast to the latter, here the global sate is not stationary. Finally, we discuss the assumptions leading to the conclusion of a stationary state in the Page-Wootters approach and in the case of Dirac's quantisation of a Hamiltonian constraint, and suggest that they need to be reconsidered.  

\section{Quantum evolution without absolute time} \label{Sec2}

In classical mechanics, time-reparameterisation invariance of a theory does not imply trivial time evolution of the physically observable instantaneous degrees of freedom (unless the value of the time parameter is assumed physically observable; see Sec. \ref{SecDisc}). Instead, for a given initial condition, the equations of motion generally yield a set of possible solutions, which are related to each other via time reparameterisation. Different solutions related in this way are considered physically equivalent, i.e.,  time reparameterisation is a gauge transformation on the configurations over time. Equivalent solutions describe the same trajectory in state space, with the difference between them being how the trajectory is traversed as a function of the time parameter, which is not determined by the equations of motion. This picture assumes an ordered continuum of instantaneous variables, or events, at the ontological level (more fundamentally, from a field theory perspective, everything is built out of spacetime point-like events, but our focus here is on time, so we consider a foliation of the spatiotemporal continuum of events into time slices). These events can be thought of as taking place over the points of a 1-dimensional time manifold, but the theory does not prescribe how to associate them with the points of such a manifold as it only prescribes relations between the events and not their relation to an external structure. Operationally, this means that we assume we could identify the variables of the system that are associated with individual instants, as well as the order between instants, but there is no separate variable at those instants that offers a parameterisation of the instants \cite{footnote5}. 

One may ask whether we could similarly assume the notion of an ordered continuum of instantaneous quantum states and potential measurements, and write an evolution equation in quantum theory, which for a given initial condition predicts the trajectory $|\psi(t)\rangle$ in the Hilbert space that is predicted by the Schrödinger equation, but does not predict a unique rate at which the trajectory is traversed and hence its exact parameterisation in terms of $t$. The intuition is that, if we assume that the Schrödinger equation correctly predicts the trajectory as a function of a time parameter measured by a specific type of external clock, if we want an evolution law that does not prescribe a relation to an external clock, the parameter along the trajectory should be left undetermined, but the trajectory as a sequence of instantaneous states should remain the same. This is because we do not assume a lack of reference frames for observables on the Hilbert space, but only a lack of a reference frame for the time parameter associated with the different instants at which such observables could be measured  \cite{footnote2}.

It is easy to construct an equation with these properties noticing that the vector $\partial_{t} |\psi(t)\rangle$ on the left-hand side of the Schrödinger equation \eqref{Schrodinger} is the tangent velocity at a given point of the trajectory. When the velocity is nonzero, the family of unit vectors pointing along the velocity at each point fully defines the trajectory, while the norm of the velocity defines the speed at a given point. We can therefore remove the determination of the speed from the equation by dividing each side by the norm of the corresponding vector. Renaming the time parameter to $\tau$ in order to distinguish it from the time $t$ appearing in the Schrödinger equation, this yields: 
\begin{subequations}
	\label{equation:main}
\begin{align}
\frac{\partial_{\tau} |\psi(\tau)\rangle} { \left|\left|\partial_{\tau} |\psi(\tau)\rangle \right|\right| }  &= -i \frac{ \hat{H}|\psi(\tau)\rangle  }{\left|\left|\hat{H}|\psi(\tau)\rangle\right|\right|}, \hspace{0.2cm} &\textrm{for} \hspace{0.2cm} \hat{H}|\psi(\tau)\rangle\neq 0, \label{equation:one}
\end{align}
where $|||\phi \rangle || =  \sqrt{\langle \phi|\phi\rangle}$. This is our new equation for nontrivial pure-state trajectories. (Note that here we consider trajectories in the Hilbert space, which is the domain of the pure-state Schrödinger equation, and not in the projective Hilbert space. Thus, we regard states differing by a global phase as different points, even though they are physically indistinguishable. Below, we will write an analogous equations for density matrices, where global phase differences are projected out. Note that, if we view this as an equation for vectors without a prior restriction on the norm, vectors differing by a real factor are also different points. But we will see that the equation preserves the norm, which allows us without loss of generality to also think of it as defined on normalised vectors.) The probabilistic interpretation of the state $|\psi(\tau)\rangle$ is assumed the same as in standard quantum theory.

In the case of a trivial trajectory predicted by the Schrödinger equation, $|\psi(t)\rangle = |\psi\rangle$, $\forall t $, which occurs when $\hat{H}| \psi(0) \rangle = 0$ for an initial condition $| \psi(0) \rangle$, the solution is trivially invariant under time reparameterisation. We can therefore extend the predictions of Eq. \eqref{equation:one} to this case as
\begin{align}
\partial_{\tau} |\psi(\tau)\rangle  &= 0,    &\textrm{for} \hspace{0.2cm}\hat{H}|\psi(\tau)\rangle= 0. \label{equation:two}
\end{align}
\end{subequations}

Eqs. \eqref{equation:one} and \eqref{equation:two}, which together define our new law of dynamics for pure states, are invariant under orientation-preserving diffeomorphisms of $\tau$. Indeed, consider such a diffeomorphism, which is given by a continuously differentiable function $\tau(\tau')$, such that $\partial_{\tau'}\tau(\tau') > 0$. Assume that $|\psi (\tau)\rangle$ respects Eq. \eqref{equation:one} with respect to $\tau$. Then, $|\psi (\tau')\rangle \equiv |\psi (\tau(\tau'))\rangle$ respects the analogue of this equation with respect to $\tau'$, because $\partial_{\tau'} |\psi(\tau')\rangle =   \partial_{\tau} |\psi(\tau(\tau'))\rangle  {({\partial_{\tau'}\tau(\tau')})}$, and hence $\frac{\partial_{\tau'} |\psi(\tau')\rangle} { \left|\left|\partial_{\tau'} |\psi(\tau')\rangle \right|\right| } = \frac{\partial_{\tau} |\psi(\tau(\tau'))\rangle \cancel{\partial_{\tau'}\tau(\tau')}} { \left|\left|\partial_{\tau (\tau')} |\psi(\tau)\rangle \right|\right| \cancel{\partial_{\tau'}\tau(\tau')}} = -i \frac{ \hat{H}|\psi(\tau (\tau'))\rangle  }{\left|\left|\hat{H}|\psi(\tau(\tau'))\rangle\right|\right|}\equiv -i \frac{ \hat{H}|\psi(\tau')\rangle  }{\left|\left|\hat{H}|\psi(\tau')\rangle\right|\right|}$, for $\hat{H}|\psi(\tau')\rangle \neq 0$. Similarly, for Eq. \eqref{equation:two} we have $\partial_{\tau'} |\psi(\tau')\rangle  =  \partial_{\tau'}\tau(\tau') \partial_{\tau} |\psi(\tau(\tau'))\rangle = 0$, for $\hat{H}|\psi(\tau')\rangle = 0$.

Furthermore, notice that if $|\psi(t)\rangle$ respects the Schrödinger equation \eqref{Schrodinger}, it automatically respects the analogue of Eqs. \eqref{equation:main} with respect to $t$. Therefore, a solution of the Schrödinger equation is always a solution of Eqs. \eqref{equation:main}, and all time-reparameterised solutions of the Schrödinger equations are also solutions. It is not difficult to see that these are in fact all the solutions of these new equations. To this end, given a specific solution $|\psi(\tau)\rangle$, in the case  $\hat{H}|\psi(\tau)\rangle\neq 0$ define $t(\tau)$ such that 
\begin{gather}
\partial_{\tau} t(\tau) =	\frac{		\left\|\partial_\tau|\psi(\tau)\rangle\right\|}{ \left|\left| 	\hat{H}|\psi(\tau)\rangle\right|\right|}.
\end{gather}
 Eq. \eqref{equation:one} then can be rewritten as
 \begin{gather}
 	\partial_{\tau} |\psi(\tau (t))\rangle = -i(\partial_{\tau} t(\tau))\hat{H} |\psi(\tau(t))\rangle,
 \end{gather}
which is equivalent to the Schrödinger equation \eqref{Schrodinger} with respect to $t$ for $|\psi(t)\rangle \equiv|\psi(\tau (t))\rangle$. In the case $\hat{H}|\psi(\tau)\rangle = 0$,  $|\psi(\tau)\rangle$ respects Eq. \eqref{equation:two}, and hence directly solves the Schrödinger equation for $t=\tau$ (or any other $t(\tau)$). Therefore, any $|\psi(\tau)\rangle$ that solves Eqs. \eqref{equation:main} is a time-reparameterised version of a $|\psi(t)\rangle$ that solves the Schrödinger equation. In particular, by invoking ancillary systems, we can conclude that the equations describe unitary evolution, albeit at an undetermined rate \cite{footnote3}. We will refer to the parameterisation $|\psi(t)\rangle$ that solves the Schrödinger equation for a specific definition of the Hamiltonian as the `Schrödinger gauge'. (Note that according to Eqs. \eqref{equation:main}, Hamiltonian operators differing by an overall, possibly time-dependent, factor are indistinguishable in the context of these equations.)

A density matrix version of Eqs. \eqref{equation:main} can be formulated analogously: 

\begin{subequations}
	\label{mixed:main}
\begin{align}
\frac{	\partial_{\tau} \hat{\rho}(\tau) }{ \left|\left|\partial_{\tau} \hat{\rho}(\tau)\right|\right|_1 }&= \frac{-i [\hat{H}, \hat{\rho}(\tau)] } {\left|\left| [\hat{H}, \hat{\rho}(\tau)]  \right|\right|_1 },&\hspace{0.2cm} \textrm{for} \hspace{0.2cm}  [\hat{H}, \hat{\rho}(\tau)] \neq 0,\label{mixed:one}\\
		\partial_{\tau} \hat{\rho}(\tau) &= 0,&\textrm{for} \hspace{0.2cm}  [\hat{H}, \hat{\rho}(\tau)]  = 0, \label{mixed:two}
\end{align}
\end{subequations}
where $\left|\left| \hat{O} \right| \right|_1 = \tr \sqrt{\hat{O}^{\dagger}\hat{O} }$ is the trace norm of $\hat{O}$. These  equations enjoy analogous properties. Note that in the case where the state is pure, $\rho(\tau) = |\psi(\tau)\rangle\langle \psi(\tau)|$, the solutions of Eqs. \eqref{mixed:main} follow from those of Eqs. \eqref{equation:main}, but the equations for $\rho(\tau) = |\psi(\tau)\rangle\langle \psi(\tau)|$ that one derives from \eqref{equation:main} are different from \eqref{mixed:main} ``off shell'', meaning that the functions appearing in them are not identical for a $|\psi(\tau)\rangle$ that is not a solution. Nevertheless, they become identical ``on shell'', i.e., for a $|\psi(\tau)\rangle$ that is a solution. 

%Finally, following the same type of construction, we can formulate an underdetermined evolution law for bounded observables $\hat{O}(\tau)$ in the Heisenberg picture. It reads: 
%\begin{align}
%	\frac{	d_{\tau} \hat{O}(\tau) }{ \left|\left|d_{\tau} \hat{O}(\tau)\right|\right|_1 }&= \frac{i [\hat{H}, \hat{O}(\tau)] + \partial_{\tau}\hat{O}(\tau)} {\left|\left| i [\hat{H}, \hat{O}(\tau)] + \partial_{\tau}\hat{O}(\tau) \right|\right|_1 },\notag\\&\hspace{0.2cm} \textrm{for} \hspace{0.2cm}  i [\hat{H}, \hat{O}(\tau)] + \partial_{\tau}\hat{O}(\tau)\neq 0,\label{mixed:one}\\
%	d_{\tau} \hat{\rho}(\tau) &= 0,\notag\\&\textrm{for} \hspace{0.2cm}  i [\hat{H}, \hat{O}(\tau)] + \partial_{\tau}\hat{O}(\tau)= 0, \label{mixed:two}
%\end{align}

Let us comment on the meaning of these evolution equations and their solutions. The parameter $\tau$ can be thought of as the time indicated by a fictitious external clock, the relation to which is unspecified by the theory. Thus, each time-parameterised trajectory $|\psi(\tau)\rangle$, or $\hat{\rho}(\tau)$, represents a possible description of the evolution relative to such a clock. As we have no physical access to such a clock, a time-parameterised trajectory is not fully physically meaningful, but the trajectory in state space that it describes is. The latter could be reconstructed in principle by the statistics of sequential measurements performed at times whose exact parameter $\tau$ is unknown, except for the fact that  it has to respect the order of the measurements---a condition that is invariant under monotonic time diffeomorphisms. Regarding the initial conditions, they are by definition also assumed given at some value of $\tau$ that in general we cannot access. But we can choose to define $\tau$ as having particular values at the times at which specific operations, such as a state preparation or a measurement, are made. In this case, the remaining gauge freedom is that of time diffeomorphisms that keep the values of the parameter at these instants fixed.

A natural question is, if these underdetermined equations provide the fundamental law of quantum evolution in a universe without an external time parameter, how can we understand the definite time evolution predicted by the Schrödinger equation? As in the Page-Wootters argument, since the Schrödinger equation holds in practice relative to physical clocks, it only needs to arise relationally. 

It is straightforward to model how such a relational description comes about in the present framework. Assume the same type of Hamiltonian \eqref{PWHam} for a system and a clock as in the Page-Wootters model. An initial state in which the clock approximately shows a definite reading, such as $|\psi(\tau=0)\rangle_{CS}\approx|t=0\rangle\otimes |\psi_0\rangle $, would evolve in the Schrödinger gauge as
\begin{gather} 
	|\psi(t)\rangle_{CS} \approx  e^{-i\hat{H}_C t} |t=0\rangle_C \otimes  e^{-i\hat{H}_S t} |\psi_0\rangle_S \notag \\= |t\rangle_C \otimes e^{-i\hat{H}_S t} |\psi_0\rangle_S,
\end{gather}
i.e., the gauge-free solution is 
\begin{gather} 
|\psi(\tau)\rangle_{CS} \approx  e^{-i\hat{H}_C t(\tau)} |t=0\rangle_C \otimes  e^{-i\hat{H}_S t(\tau)} |\psi_0\rangle_S \notag \\= |t(\tau)\rangle_C \otimes e^{-i\hat{H}_S t(\tau)} |\psi_0\rangle_S,
\end{gather}
where $t(\tau) $ is an arbitrary monotonic diffeomorphism of $\tau$. Here, the clock and the system evolve with $\tau$ synchronously as if following the Schrödinger equation, but at an undetermined rate. When the state of the clock is approximately $| t \rangle_C$, the state of the system is approximately $e^{-i\hat{H}t}|\psi(0)\rangle$, i.e., the state of the system follows the Schrödinger equation relative to the reading of the clock. Unlike the Page-Wootters model, however, here the global state is not stationary and in principle could be subjected to measurements in agreement with the standard rules of quantum theory. For instance, an agent could perform a measurement of the internal clock $C$ in the $|t\rangle_C$ basis at an unknown instant $\tau$, and they would find a specific value $t$, causing negligible disturbance to $C$ similarly to the measurement of a classical clock in everyday situations. They could also make a measurement on $|\psi(t)\rangle_S$ at the same instant, which would yield an outcome and change the state according to the standard rules of quantum theory, affecting the evolution in the future both with respect to the internal clock and the external clock, the relation between which remains unspecified by the theory. The agent could also make more general measurements on the joint system $CS$ at a given $\tau$, which can be treated by the same standard rules. 

\section{Discussion} \label{SecDisc}

We have seen that it is possible to formulate time-reparameterisation invariant evolution equations in quantum theory, which predict nontrivial trajectories of the global state in agreement with the Schrödinger equation, but do not predict the speed of evolution along these trajectories. This is to be contrasted with the stationary global state picture in the Page-Wootters model and in Dirac's quantisation. We now examine in turn the assumptions  leading to a stationary state in these formalisms. 

The Page-Wootters formalism \cite{PW} assumes a superselection rule for energy at the level of the full universe, which the authors motivate by two separate considerations. On the one hand, in analogy with the superselection for electric charge, which arises from the fact that the total charge operator commutes with all quasilocal observables as a result of the long-range Coulomb field \cite{Chargesuperselection}, the authors suggest that one may expect a similar property to hold for energy, since energy couples to a long-range gravitational field. On the other hand, they argue that even in the absence of gravity, the unobservability of a global time shift for the full universe implies that only operators that have no dependence on external time can be observables, which gives rise to an energy superselection rule for the total Hamiltonian.

Regarding the first consideration, it is not clear if the analogy with electric charge holds, because the derivation of the superselection rule for electric charge assumes a background Minkowski spacetime \cite{Chargesuperselection}, whereas in general relativity there is no background spacetime. Since general relativity is invariant under spacetime diffeomorphisms, which include time reparameterisations, its Hamiltonian for a closed universe is proportional to a constraint, the logic of whose quantisation will be discussed below. (In an open universe with fixed boundary conditions, such as an asymptotically flat spacetime, the boundary term of the Hamiltonian is generally not expected to commute with dressed bulk observables, which extend to the boundary \cite{DonnellyGiddings}. But irrespectively of whether one can conclude a superselection of the boundary energy for a certain notion of bulk observables, below we question the very treatment of the Hamiltonian constraint in the bulk.)

The second, more general, consideration argues that the unobservability of a global time shift implies that physical observables should commute with the Hamiltonian. As commonly understood, this argument assumes a lack of a reference frame for global time translations \cite{Loveridge}, relative to which the system would follow standard evolution, since it emphasises the impossibility of detecting a global time shift, based on which it concludes that accessible observables must commute with the Hamiltonian. We will thus perform our analysis under these assumptions, but the essential conclusions extend to the more general symmetry of time reparameterisation. %Below we also comment on another possible, more radical, assumption---inability to identify the order of instants. 

In principle, when an agent loses access to reference frames for the parameters associated with a certain group of transformations, the observables that remain accessible to them are those that are invariant under the action of the group. See Ref. \cite{extraparticle} for a proof of this fact using the theory of quantum reference frames for symmetries that do not involve time. Now, it is not immediate to apply this principle to the case of time translation, because time translation is not simply a transformation on individual operators on the Hilbert space---it acts on families of operators parameterised by time (and possibly more general objects constructed out of these families, such as probabilistic mixtures of families, as discussed in the \hyperref[sec:Appendix]{Appendix}). Examples of such families are a solution $\hat{\rho}(t)$ of the Schrödinger equation, a time-dependent observable $\hat{O}(t)$ in the Heisenberg picture (in this paper, we do not develop underdetermined evolution equations in the Heisenberg picture, but this is an interesting question for future investigation), or a sequence of transformations $\hat{M}(t)$ applied on the system at different times. Assuming that the time parameter $t$ ranges in $(-\infty, \infty)$, a time translation by an amount $\Delta t$ maps a family of operators $\hat{O}(t)$ to the family $\hat{O}(t+\Delta t)$. The family is invariant under arbitrary time translations, if and only if $\hat{O}(t) = \hat{O}(t’)$, $\forall t, t’$. Now, if we consider an observable in the Heisenberg picture that corresponds to a time-independent observable in the Schrödinger picture, it satisfies $\hat{O}(t+\Delta t) = e^{i\Delta t\hat{H} }\hat{O}(t) e^{-i\Delta t\hat{H} } $. Therefore, the family describing such an observable is invariant under time translations if and only if $\hat{O}(t)= \hat{O}$, $\forall t$, where $[\hat{O},\hat{H}]=0$. One may thus be led to conclude that the only physically meaningful static observables in the Schrödinger picture in the absence of a reference frame for time translations are observables that commute with the Hamiltonian, which effectively imposes the stated superselection rule. 

However, this reasoning assumes that a solution of the equations of motion is given by a unique family of states in the Schrödinger picture, or by a unique family of observables for a given observable in the Heisenberg picture. But as we have seen, a solution of the evolution equations need not be described by a single family. In our approach, a physically distinct solution in the Schrödinger picture corresponds to a \textit{set} of possible families of states, where the different families are related to each other via transformations from the symmetry group, and are viewed as physically equivalent. This set is invariant under the action of the group since applying an element of the group on all members of the set maps the set to itself. In the case of the full reparameterisation group, the invariant degrees of freedom that are common to all members of the set and uniquely define the set are given by the trajectory in state space, which is generally nontrivial. If we lack a reference frame only for time translations, we can similarly define an equivalence class of solutions that are obtained from a given solution of the Schrödinger equation by applying to it all possible time translations, and such a class in general would be characterised by a nontrivial trajectory. One may nevertheless wonder whether the fact that we cannot physically distinguish the different members of the class is not operationally equivalent to having a trivial trajectory in state space. In the \hyperref[sec:Appendix]{Appendix}, we explain why this is not so. In particular, we show that while for a single measurement the predictions are indistinguishable from those of the Page-Wootters picture, the situation changes radically for sequential measurements, which can be performed in a time-translation invariant way. There, nontrivial global time evolution can be observed through the correlations between the measurements, as is done when we have a background time parameter. The situation is qualitatively similar for time-reparameterisation invariance.

Therefore, the Page-Wootters conclusion requires more than just inability to observe an external time parameter. It may be possible to derive it assuming inability to identify the order of instants, or to recognise instants altogether. However, this is strictly stronger than the gauge freedom know to be physically relevant from general relativity. 

One may ask whether the symmetries of time translation and reparameterisation that we  have discussed here, and the corresponding invariant objects, could be studied using the tools of group representations over a Hilbert space. In Ref. \cite{BouvyOreshkov}, we present a formalism that describes quantum evolutions by operators on a suitable large Hilbert space, which offers such possibilities.

Finally, we come to the question of Dirac's quantisation of the Hamiltonian constraint. The projection on the subspace annihilated by first-class constraints in Dirac's approach can be seen to successfully isolate the gauge-invariant degrees of freedom when the constraints are generators of gauge transformations in the phase space. However, the latter is true when the time parameter appearing in the theory is assumed physically observable \cite{Henneaux}, as in a gauge theory such as electromagnetism. In that case, the constraints, which appear as terms in the Hamiltonian with unspecified time-dependent prefactors, generate arbitrary changes in the state at a given time that can only be gauge transformation if the evolution equations must prescribe a unique physical solution. However, when the time parameter itself is the gauge variable, viewing the Hamiltonian constraint as a generator of gauge transformations in this sense is no longer justified. Indeed, there is no reason to think that the phase-space parameters that get changed by the action of the Hamiltonian as a generator of transformations in that space are physically unobservable. What is unobservable is the time at which that change happens. This raises doubts as to whether imposing Eq. \eqref{DiracHamConstraint} is a correct quantisation procedure when the unphysical degree of freedom is the time parameter itself. Rather, this equation seems to correspond to the quantisation of a time-reparameterisation invariant classical theory in which the time parameter is assumed physically observable. In that case, also at the classical level all phase-space transformations generated by the Hamiltonian are gauge transformations and there is no physical evolution. A similar observation suggesting the need to rethink canonical quantisation in view of the so-called local interpretation of general relativity has been made in Ref. \cite{RovelliQRS}.
	
In conclusion, our results point out an alternative approach to realising time-reparameterisation invarinace in quantum theory---by means of underdetermined equations---and suggest reconsidering the canonical methods of quantising classical theories with this symmetry. This may have nontrivial implications for the theory of quantum gravity.

\textit{Acknowledgements.} O. O. thanks Esteban Castro-Ruiz and Carlo Rovelli for stimulating discussions. This work was supported by the F.R.S.-FNRS under project CHEQS within the Excellence of Science (EOS) program. This publications was made possible through the support of the ID\# 63683 grant from the John Templeton Foundation, as part of the \href{https://www.withoutspacetime.org}{‘WithOut SpaceTime’ Project (WOST)}. The opinions expressed in this publication are those of the authors and do not necessarily reflect the views of the John Templeton Foundation.  O. O. is a Senior Research Associate of the Fonds de la Recherche Scientifique (F.R.S.-FNRS).

\section*{Appendix}\label{sec:Appendix}

To gain further understanding of the operational implications of lacking access to an external reference frame for time translations, imagine that we perform a measurement at some unknown time $t$ according to that reference frame. If no value of $t$ is more likely than another, the operation is invariant under time translations and should in principle be possible to implement without access to the external frame. In the Heisenberg picture relative to the external frame, the measurement operator corresponding to a given outcome of that measurement would be one out of a set of possible operators of the form $\hat{E}(t) = e^{i t\hat{H} }\hat{E} e^{-t\hat{H}} $ over all possible $t \in (-\infty, \infty)$. We would now like to formalise the idea that all values of $t$ according to an agent with access to the external reference frame are `equally likely', and to define an averaged positive operator-valued measure (POVM) element corresponding to that outcome, which would yield the probability of that outcome in agreement with the Born rule for any Heisenberg state that may describe the information of the external agent (we can think that this state is defined at $t\rightarrow -\infty$ so that a measurement with its corresponding update rule can be applied on it at any time). 

If the parameter $t$ took values in a compact set, we could define a uniform probability density over it, in which case the desired POVM element would be the average of the operators $\hat{E}(t)$ for this uniform density. However, the time-translation group is not compact. Nevertheless, we can argue that the POVM element corresponding to such a random-instant measurement without any bias over the possible time translations of that instant relative to the external frame should be given by an operator of the form $\hat{\overline{E}}=\sum_i \hat{P}_i \hat{E} \hat{P}_i$, where $\{\hat{P}_i\}$ is the complete set of mutually orthogonal projectors on the eigenspaces of the Hamiltonian $\hat{H}$.  In other words, $\hat{\overline{E}}$ can be obtained from any element $\hat{E} (t)$ from the set (here, we have taken the element $\hat{E} (0) \equiv \hat{E}$) by acting on it with the completely positive trace-preserving linear projector $(\cdot) \rightarrow \sum_i \hat{P}_i (\cdot) \hat{P}_i$ on the space of operators that commute with the Hamiltonian. (Note that this projector maps the identity operator to itself, so it maps every POVM to a POVM.) 

This can be obtained as follows. Consider the set of operators $\hat{E}(t)$ over the finite interval $t \in (-T,T)$, $T>0$, and define the average operator for a uniform distribution over this interval, $\hat{\overline{E}}_{T} \equiv \frac{1}{T}\int_{-T/2}^{T/2} d t \hat{E}(t) = \frac{1}{T}\int_{-T/2}^{T/2} dt   e^{i t\hat{H} }\hat{E} e^{-t\hat{H}}$. This operator has the property that $\hat{P}_i\hat{\overline{E}}_{T} \hat{P}_i = \hat{P}_i\hat{E} \hat{P}_i$, $\forall i$. Furthermore, in the limit $T\rightarrow \infty$, the off-diagonal blocks of $\hat{\overline{E}}_{T}$ vanish, i.e., $\lim_{T\rightarrow \infty} \hat{P}_i \hat{\overline{E}}_{T} \hat{P}_j = 0$, $\forall i \neq j$, because these terms are proportional to $\frac{1}{T}\int_{-T/2}^{T/2} d t e^{i t (h_i-h_j)}\underset{T\rightarrow \infty}{\rightarrow} 0$, where $h_i$ are the Hamiltonian eigenvalues. In other words, we obtain $\hat{\overline{E}}=\sum_i \hat{P}_i \hat{E} \hat{P}_i = \lim_{T\rightarrow \infty} \hat{\overline{E}}_{T}$ as the limit of a well defined average. This limit is invariant under time translations, in the sense that the result is the same if we shift all values of the time parameters in this derivation by any finite amount, as expected from an object describing an operation without access to the external reference frame. 

At this point, we seem to have recovered the Page-Wootters conclusion that only measurements that commute with the Hamiltonian can be performed on the system, which yields a superselection rule preventing global evolution from being observed. However, this analysis considers only a single measurement performed on the system. For such a measurement, the result is indeed indistinguishable from the result of performing a measurement whose measurement operators commute with the Hamiltonian. But if we consider multiple measurements performed sequentially, the situation changes drastically.  If after a measurement of the type above we perform another measurement exactly time $\Delta t$ later (this is operationally meaningful since the distance in time between two events is invariant under time translation), we would observe correlations in agreement with the property that the evolution between the two instants follows the Schrödinger equation in $t$. For example, if the universe contains a clock as in the model from the end of Section \ref{Sec2}, two sequential measurements on it in the `clock basis' (recall that for a realistic clock these cannot be ideal projective measurements of a continuous parameter, so they are generalised measurements or projective measurement that show discrete values) would show with high probability a difference of approximately $\Delta t$. This strong correlation occurs despite the fact that a single such measurement would give a completely random result. Indeed, the effective POVM we would obtain for such a measurement by `averaging' over all time translations is trivial, with all outcomes proportional to the identity operator, so the outcomes of sequential independent measurements of this kind could not be correlated. This highlights the fact that the heuristic `uniform distribution' on which this averaging is based is a distribution of {time-parameterised families of operators}, which describe \textit{sequences} of events. This `distribution' captures nontrivial evolution in the correlations between different instants, despite being time-translation invariant. If we focus on a single instant, we see average operators that commute with the Hamiltonian as in the Page-Wootters model, but this misses the correlations in the bigger space of operators over time.

To summarise, the lack of a reference frame for time translations by itself does not imply impossibility of detecting time evolution. It only implies impossibility of measuring the time difference between observed events and an unobservable external event that defines the origin of our time coordinate. The situation is similar in the case of a lack of a reference frame for time reparametrisations, where we would further be unable to measure the time difference between observed events, but could still know their order.


\begin{thebibliography}{1}

\bibitem{PW} D. N. Page and W. K. Wootters,  Evolution without evolution: Dynamics described by stationary observables, \textit{Phys. Rev. D.} \textbf{27}, 2885 (1983).

\bibitem{footnote1} There is no Hamiltonian bounded from below that can generate this ideal translation for a continuous orthonormal basis $|t\rangle$ \cite{Pauli}. In practice, a realistic clock can only keep track of time approximately since its normalised, physical states $|s(t)\rangle$, $\langle s(t)|s(t)\rangle = 1$, $\forall t$, cannot be strictly orthogonal to each other for different $t$, $\langle s(t)|s(t')\rangle \neq 0$, $\forall t \neq t'$, and their precision in indicating time via a fixed observable would tend to decrease with time. However, for any finite precision and finite time interval, it is in principle possible to construct a physical clock that works with that precision during the interval.

\bibitem{Pauli} W. Pauli, General Principles of Quantum Mechanics (\textit{Springer-Verlag}, 1980). Original article published in German in 1933. 

\bibitem{Dirac} P. A. M. Dirac, Lectures on Quantum Mechanics, published by Belfer Graduate School of Science, Yeshiva University, 1964; (\textit{Dover Publications}, 2001). 

\bibitem{DeWitt} B. S. DeWitt, Quantum Theory of Gravity. I. The Canonical Theory, \textit{Physical Review} \textbf{160}, 1113 (1967). 

\bibitem{Henneaux} M. Henneaux and C. Teitelboim, Quantization of Gauge Systems, (\textit{Princeton University Press}, 1992).

\bibitem{Giovanetti} V. Giovanetti, S. Lloyd, and L. Maccone, Quantum Time, \textit{Phys. Rev. D} \textbf{92}, 045033 (2015). 

\bibitem{Colosi} D. Colosi and C. Rovelli, A simple background-independent hamiltonian quantum model, \textit{Phys.Rev. D} \textbf{68}, 104008 (2003). 

\bibitem{footnote5} Consider the following abstraction. Imagine that we have a continuous (and hence uncountable) set of photos stacked horizontally from left to right, where between any two photos there are infinitely many others. We could pick up and look at any photo, and we could know the order of two photos based on whether one is on the left or on the right of the other, but nothing more. In such a case, if we have to deterministically assign a real parameter on the photos, we have no way of doing this except based on the content of the photos, which is analogous to fixing a gauge. If this has to be done independently of the content of the photos, there would need to be a separate variable associated with each photo (like its position relative to a background, or its distance from another photo), and this is what we assume we have no access to. 

\bibitem{footnote2} One may investigate how to describe a universe in the absence of various types of reference frames. A methodology that has proven useful for understanding individual symmetries is to imagine that reference frames for other degrees of freedom are available and relax only the assumption of reference frames for the degrees of freedom of interest \cite{Poulin, BRS, extraparticle}. Since here we are interested in time reparameterisation, we only assume the absence of an external time parameter. 

%Note that it has been suggested \cite{??} in the context of certain thought experiments involving events with indefinite causal order and quantum clocks in superposition that the notion of space-time locality itself should be considered relative. This potential extension of the present framework is left for future investigation. 


\bibitem{Poulin} D. Poulin, Toy Model for a Relational Formulation of Quantum Theory, 
\textit{Int. J. Theor. Phys.} \textbf{45}, 1189 (2006).

\bibitem{BRS} S. D. Bartlett, T. Rudolph, and R. W. Spekkens, Reference frames, superselection rules, and quantum information, \textit{Rev. Mod. Phys.} \textbf{79}, 555 (2007)/

\bibitem{extraparticle} E. Castro-Ruiz and O. Oreshkov, Relative subsystems and quantum reference frame transformations, \textit{Commun. Phys.} \textbf{8}, 187 (2025)


\bibitem{footnote3} In order to make sense of the claim that the evolution relative to $\tau$ is unitary, we need to be able to say how different possible initial states would evolve during the same interval as measured by $\tau$. At first sight this may look like a meaningless notion since the parameter $\tau$ is gauge. However, a concrete instant is still presumed an operationally meaningful notion, and there is a way of defining how different initial states would evolve up to the same instant by appealing to the evolution of a larger system that includes an ancillary system. Imagine that in addition to the system that we are describing, call it $Z$, there is an ancilla $A$ whose Hamiltonian vanishes. If the initial state on the joint system is a pure entangled state $|\psi(0)\rangle_{ZA}$ such that the reduced density matrix on the system, $\hat{\rho}_Z(0) = \tr_A(|\psi(0)\rangle\langle \psi(0)|_{ZA})$, has full support, then by making a suitable measurement on the ancilla so as to project it on a specific state, we can, with nonzero probability, obtain any desired pure state on the system. Imagine now that we let the joint system evolve until some subsequent instant, which would corresponds to some $\tau\neq 0$. We know that the joint state at this instant would be of the form $\hat{U}_Z(t(\tau)\otimes \hat{\id}_A   (|\psi(0)\rangle\langle \psi(0)|_{ZA}) \hat{U}^{\dagger}_Z(t(\tau))\otimes \hat{\id}_A $, where $\hat{U}_Z(t(\tau)$ is the unitary generated by the Schrödinger equation for some time $t(\tau)$. Since a measurement on $A$ commutes with the Hamiltonian on $Z$, any pure state on $Z$ that we would obtain at the instant $\tau$ by projecting the ancilla $A$ on a specific state is the same as the result of the evolution up to that instant applied on the initial state that we would obtain if we make the same projection at $\tau=0$. Obviously, all states that we could prepare in this way at the initial instant would be transformed to the final instant via the same unitary $\hat{U}_Z(t(\tau))$.

\bibitem{Chargesuperselection} F. Strocchi and A. S. Wightman, Proof of the Charge Superselection Rule in Local Relativistic Quantum Field Theory, \textit{Math. Phys.} \textbf{15}, 2198 (1974).

\bibitem{DonnellyGiddings} W. Donnelly and S. B. Giddings, Diffeomorphism-invariant observables and their nonlocal algebra, \textit{Phys. Rev. D} \textbf{93}, 024030 (2016).

\bibitem{Loveridge} L. Loveridge and M. Miyadera, Relative Quantum Time, \textit{Found. Phys.} \textbf{49}, 549 (2019). 

\bibitem{BouvyOreshkov} D. Bouvy and O. Oreshkov, The continuous circuit operator: a formalism for quantum theory in spacetime (in preparation); D. Bouvy, Master thesis, ULB (2026). 

\bibitem{RovelliQRS} C. Rovelli, Quantum reference systems, \textit{Class. Quantum Grav.} \textbf{8} 317 (1991). 

\end{thebibliography}
\end{document}